\documentclass[aip,prl,twocolumn,amsmath,amssymb,superscriptaddress,final,floatfix,a4paper]{revtex4}
\usepackage{epsfig}
\usepackage{epstopdf}
\usepackage{color}
\usepackage{graphicx}
\begin{document}
\title{Equilibrium  Gels of Limited Valence Colloids}

\author{Francesco Sciortino}
\affiliation{Sapienza--Universit\`a di Roma, P.le A. Moro 5, 00185 Roma, Italy}
\affiliation{ CNR-ISC, Sapienza University of Rome, Piazzale Aldo Moro 5, I-00185 Roma, Italy}
\author{Emanuela Zaccarelli}
\affiliation{ CNR-ISC, Sapienza University of Rome, Piazzale Aldo Moro 5, I-00185 Roma, Italy}


\begin{abstract}
Gels are low-packing arrested states of matter which are able to support stress.  On cooling, limited valence colloidal particles form open networks  stabilized by the progressive increase of the interparticle bond lifetime.  These gels, named {\it equilibrium gels}, are the focus of this review article. Differently from other types of colloidal gels, equilibrium gels do not require an underlying phase separation to form.  Oppositely, they form in a region of densities  deprived of thermodynamic instabilities. Limited valence equilibrium gels neither coarsen nor age with time.
\end{abstract}

\maketitle

\section{Introduction}

Gels are ubiquitous in nature.  At the atomic and molecular level,  network liquids 
form extended open transient networks of bonds of quantum-mechanics origin (tetrahedral in water and silica)~\cite{geiger1979aspects,stanley1980interpretation,epjb}.
In the amorphous  state, the bonding pattern is frozen in a permanent structure. At polymer level, gels arise from multiple chemical or physical connections between distinct chains~\cite{colby2003polymer,tanaka2011polymer}.  At colloidal level, gels occur when  the inter-particle interaction strength becomes considerably larger than the thermal energy $k_BT$ and particles stick together forming a disordered  highly porous material.

The origin of the colloidal gel state is one of the fascinating questions in soft matter~\cite{zaccareview}: 
why particles prefer to form an arrested state of matter (a disordered solid), despite the very limited amount of  occupied volume, i.e. in the absence of excluded volume caging effects? 
Does colloidal gelation always require an underlying phase-separation? When and how the gel state is a real equilibrium state of matter, that does not age or coarsen with time?  Some of these questions are discussed in this article.

We specifically focus on particle colloidal gels~\cite{zaccareview,Sciort_04,sciortino2011reversible}, e.g. on systems in which  the monomeric unit is a particle, although some of the concepts we will develop are also relevant to molecular and polymeric gels.  While we deliberately discuss only one-component gels, the same ideas can be generalized to multicomponent systems~\cite{de2012bicontinuous,lindquist2016formation}.  We thus concentrate on 
systems in which  gelation arises from the onset of a network of  long-lived interparticle bonds between identical particles.   The solvent, when present, and its quality enter only in the definition of the effective interaction potential between particles.  

In analogy with chemical gels (e.g. systems with permanent  bonds), colloidal gelation has often been identified with percolation, e.g. with the spontaneous formation of an infinite
spanning cluster that provides rigidity to the material.   However, very often, at  the percolation locus (a line in the temperature-density  plane), the lifetime of the
inter particle bonds is very short and no dynamical signatures of arrest are detected in experimental observables. Under these conditions, bonds are reversible so that the incipient spanning cluster is transient and it restructures itself on the same timescale of the bond lifetime.  Gelation in these systems thus requires temperatures significantly  smaller than the $T$ at which  percolation is
encountered. As a result, kinetic arrest in a disordered solid-like structure is commonly observed when $T$ is much smaller than the inter particle attraction strength, e.g. in the same $T$-window where
phase separation takes place. Notable exceptions, discussed below, are provided by systems
in which the bond lifetime is decoupled from the thermodynamics of the system, as in systems in which   strong activation barriers separate bonded and non-bonded states~\cite{delgado2,saikabond}. 

The most common colloidal gels (depletion gels~\cite{verhaegh,ilett}) are formed by a quench into a thermodynamic unstable region, followed by spinodal decomposition and kinetic arrest, which finally prevent a complete phase-separation~\cite{LuCiulla,piazza}. This process creates dynamically arrested structures encoding information of the underlying decomposition process in the resulting
material~\cite{leone1987order,zaccareview,chiappini}. Such gels are in a non-equilibrium state and display restructuring processes aiming at slowly completing phase-separation.  
In the case of extremely deep quenches, the phase separation  becomes equivalent to an irreversible aggregation process, commonly indicated as  diffusion limited cluster aggregation (DLCA). Under these conditions,  fractal aggregates grow, progressively filling up space until a macroscopic structure forms.  As shown in several  experiments~\cite{carpineti1992spinodal, cipelletti2000universal,LuCiulla} and theoretical studies~\cite{sciortino1995structure},  the structure of   DLCA gels displays the hallmark of spinodal decomposition, e.g. a coarsening peak in the static structure factor at low wavevectors.  
 
Another type of colloidal gels originates from the 
competition between short-range attraction (often of depletion origin) and long range repulsion (often of electrostatic origin). Such competition suppresses phase separation in favour of the formation of finite size aggregates (microphase separation). Depending on the  actual shape of the inter particle potential, these clusters can be highly anisotropic and give rise to slowly aging filamentous gels~\cite{campbell2005dynamical,toledano2009colloidal}.

In the last decade,  the concept of equilibrium colloidal gels, i.e. gels that do not age and display a remarkable thermodynamic stability,  has been developed in a series of studies  primarily aiming at understanding the phase behavior of limited-valence patchy colloidal particles~\cite{Zacca1,bian,23,lungo,krzakala2008lattice,sastry2006maximum},  establishing a strong connection between the limited valence at particle level and the formation of an equilibrium gel as a collective state.
These new colloids are characterized by highly anisotropic interactions, mimicking bonding in network liquids~\cite{epjb,biffi2015equilibrium}.  In  colloids interacting via isotropic attractive forces~\cite{lekkerkerker1992phase,anderson2002insights,lekkerkerker2011colloids,koumakis2011two} (as for the depletion interaction~\cite{asakura1954interaction})  the condensed phase (the analogue of the liquid phase in atoms) is typically constituted by a  dense local structure.  Indeed, to minimize the system energy, particles are surrounded by the  geometrically controlled maximum number ($\approx 12$) of   neighbours.      If such particles  are brought to a temperature significantly smaller than the attraction energy scale (or equivalently if the 
attraction strength is significantly larger than thermal energy at ambient temperature), particles will cluster into 
aggregates, eventually separating into a dilute gas of monomers and a dense liquid.  When the colloidal particles interact via  a limited number  of bonds (or valence $f$) the lowest energy state is achieved when all possible bonds are formed.  But now the number of 
bonded neighbours is fixed by valence.  The resulting {\it empty} 
liquid\cite{bian} is characterised by a density that is approximately reduced by a factor $f/12$ with respect to the isotropic case.  If a system of limited-valence patchy  colloids is then brought to a temperature significantly smaller than the attraction energy scale, particles will cluster into open aggregates. Only at low densities, it will eventually separate into a  gas of monomers and an  {\it empty}  liquid,
 the structure of which can be envisioned as an open network of bonded particles.  For tetrahedral patchy colloids, the typical volume fraction occupied by particles in the empty liquid state is  approximatively 25-30 $\%$, as compared to the $\approx 60 \%$ packing in simple liquids.

\begin{figure}[h]
\begin{center}
\includegraphics[width=10cm]{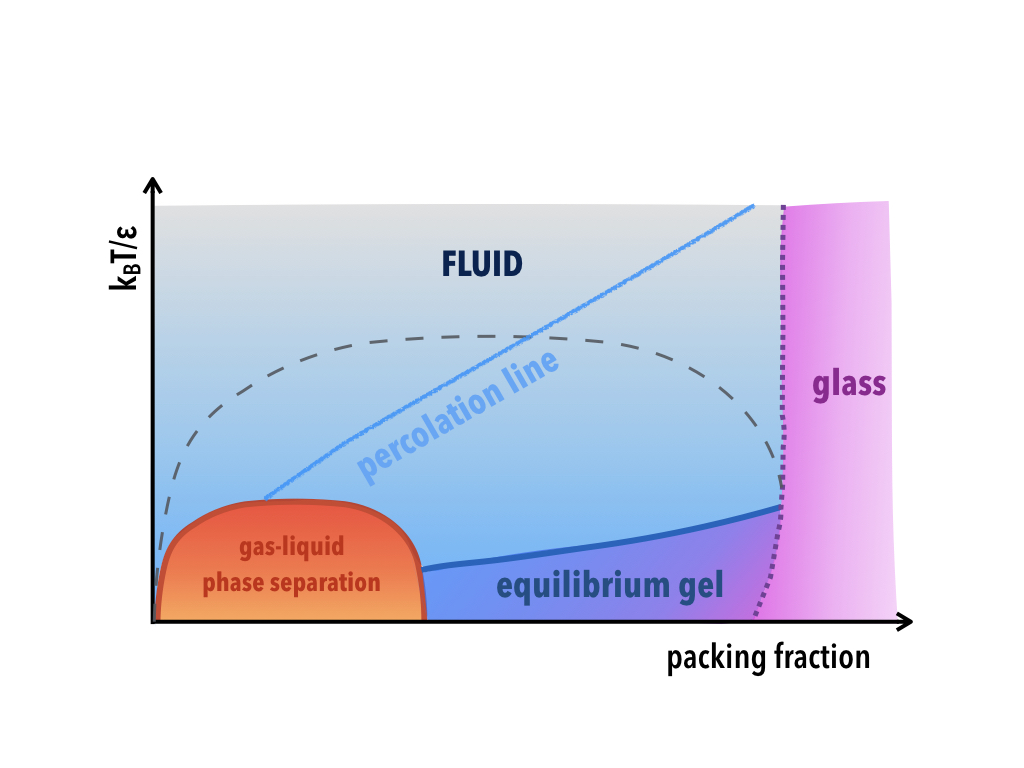}
\caption{Schematic phase diagram of limited valence particles, particles 
interacting via the excluded volume repulsive interaction complemented by a small number of
attractive sites  (patches).  Compared to the corresponding isotropic case  (e.g. same depth and same range of the attractive potential), the gas-liquid phase separation is only present  in a limited region of densities (full line: limited-valence case, dashed line: isotropic case).  
The critical temperature is also significantly reduced compared to the corresponding isotropic case.  Between the phase-separation region and the high density glass region, a window of intermediate densities exists in which the system can be brought at very low temperatures in a homogeneous state.   As for the glass case, in which arrest is driven by caging induced by excluded volume interactions, the arrest line  in this intermediate density region  
can be empirically defined as the line at which the relaxation time of the system (slaved by the bond lifetime) becomes  comparable to the experimental time scale. Below such line the system becomes arrested, forming an equilibrium gel.  Note that equilibrium gels can not be formed when particles interact via standard (core repulsion plus attraction) isotropic potentials, since in this case the gas-liquid coexistence curve extends up to the glass region.   
The percolation locus, defined as the locus at which an incipient spanning cluster of bonded particles first appears, is always located at $T$ higher than the gas-liquid critical temperature. Commonly, the bond lifetime at the percolation locus is much smaller than macroscopic observation times.
Thus, even well below percolation, the system retains its fluid state.   In  very rare cases
in which the bond lifetime is comparable or larger then the macroscopic observation time  
at the percolation locus, then gelation coincides with percolation. 
}
\label{fig:phase}
\end{center}
\end{figure}

What we are learning from these considerations is that limited valence is a fundamental requisite to
generate open stable equilibrium phases at low $T$. 
A schematic representation of the phase-diagram of low-valence particles in shown in Fig.~\ref{fig:phase}. Thus a window opens up 
in  the density region  between the gas-liquid coexistence and the glass (where excluded volume dominates)  in which the liquid state remains  stable. 
Upon cooling the system inside this density region, particles becomes progressively connected
by  physical bonds, until the system reaches its lowest energy state, in which all possible bonds are formed.  Further cooling does not change any longer the structural properties of the gel, affecting only the time-scale of bond breaking and reforming. In the most simple conditions, such dynamics is controlled by the energy scale of the bond and follows an Arrhenius law.   When the lifetime of the bonds becomes longer than the experimental time-scale,  the network structure does not change during the observation time and the system behaves as a gel.

In summary, equilibrium gels do not require an underlying phase separation. They form at low $T$ 
in  a  intermediate density window which exists only when particles primarily interact via a limite number of directional interactions.
The empty liquids progressively transform in equilibrium gels with the increase of the inter-particle bond lifetime on cooling.
Equilibrium gels are  thus non-aging percolated networks at a  low particle packing fraction, whose value is controlled by the valence.
When the average $f$ approaches two (for example by mixing together bi- and three-functional colloids), 
the system forms very long chains of bi-functional particles  with  branching points
provided by the multi-functional particles, resulting in a very empty network~\cite{bian,23,heras2011phase,de2012bicontinuous}.

\section{Experiments on limited valence colloids and equilibrium gels}

\begin{figure}[h]
\begin{center}
\includegraphics[width=8cm]{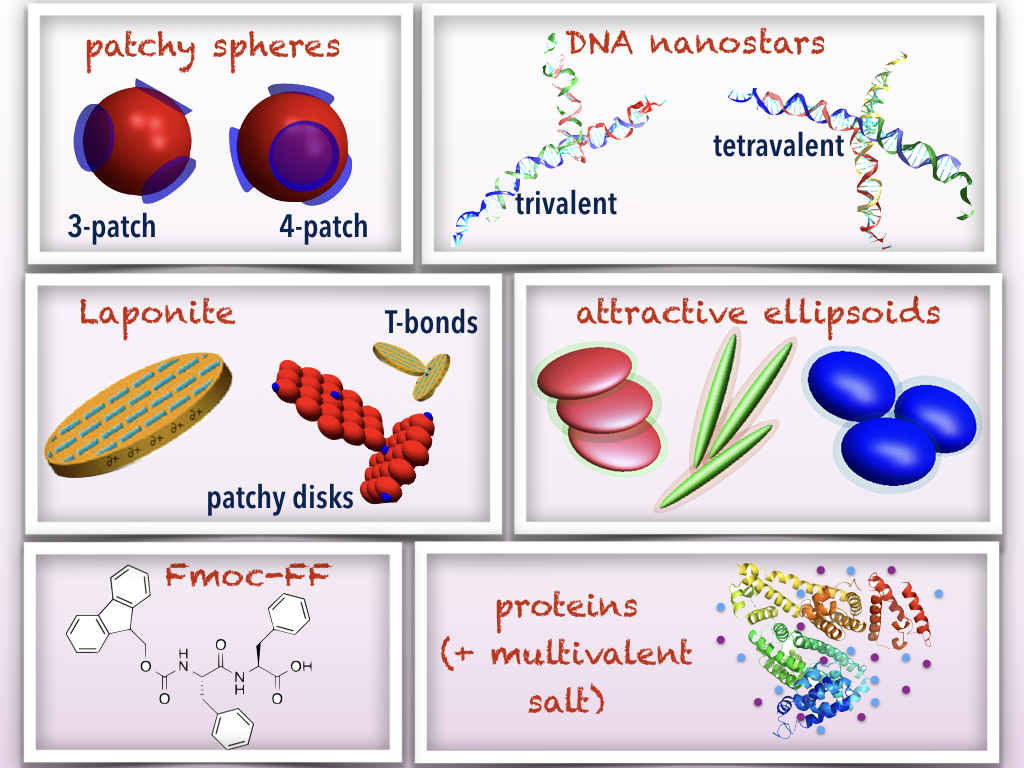}
\caption{Illustration of several model systems, both theoretical and experimental,  discussed in this article. 
In order, a patchy particle model~\cite{Kern_03}, the DNA-nanostars,  the laponite discotic platelet~\cite{ruzicka2011fresh},  oblate and prolate attractive ellipsoids~\cite{meneses2013towards},
the Fmoc-FF molecule~\cite{dudukovic2014mechanical} and a protein in the presence of multivalent salt providing limited interactions~\cite{zhang2012charge}.}
\label{fig:example}
\end{center}
\end{figure}

In the last years a significant effort has gone in the direction of synthesizing colloidal
particles with anisotropic interactions~\cite{Glotz_Solomon_natmat,van2003chemistry}. 
Beyond Janus colloids~\cite{perro2005design,jiang2010janus}, particles  whose surface is divided in two part with different
physico-chemical properties,  particles with multiple patches are becoming
progressively more  available~\cite{jiang2009simple,kegel,chen2011directed,chen2011triblock,kraft2012surface,
wang2012colloids,Paw10a,glad1,kamalasanan2011patchy,yi2013recent,sacanna2013shaping,duguet2016patchy}.  The possibility to modify the patch-patch interaction acting for example on the hydrophobicity of the surface  or by binding 
on the patch surface specific chemical species (often a DNA sequence to capitalize on the
bonding selectivity) opens the way to a very large number of future applications and experimental
realizations of equilibrium colloidal gels.  
Fig.~\ref{fig:example} shows a cartoon of the model systems which have been so-far investigated to
highlight the equilibrium gel behaviour, discussed in the following sections.

\subsection{Laponite}

The first experimental evidence of a phase-coexistence limited to very small densities, contiguous to a gel state~\cite{ruzicka2011observation}
has been provided by a study of Laponite, an industrial synthetic clay that, dispersed in water, forms a suspension of nanometre-sized discotic platelets with
inhomogeneous charge distribution and directional interactions~\cite{ruzicka2011fresh}(Fig.~\ref{fig:example}). It has been suggested that
the opposite sign of the charges on the rim and on the face of each platelet favours the formation
of T-configurations in which the rim of one platelet sticks to the face of a neighbour one.
The limited number of such T-configurations is responsible for the limited number of contacts (valence).
The strength of this electrostatic bond is significantly larger than the thermal energy, de facto
giving rise to an irreversible kinetics in which bonds are progressively formed.
Experiments show that  for concentration smaller than $1\%$ wt  the samples undergo an extremely slow phase-separation process into a clay-rich  (gel) and a clay-poor (fluid) phases. 
 This very slow phase separation process involves a transient in which the system, which forms a 
 fluid solution when prepared becomes progressively solid-like before the final macroscopic  separation  takes place.
Independently from
the starting concentration (but smaller than $1\%$ wt)  the final concentration of the
clay-rich phase is always $1\%$ wt.  Samples prepared with concentration 
larger than  $1\%$ wt  remain in a homogeneous state at all times. 
 X-rays scattering experiments~\cite{ruzicka2011observation}  have confirmed
that for $c<1\%$ wt the scattering intensity progressively increases with waiting time, a signature of a very
slow ongoing phase-separation, while for $c>1\%$ wt it approaches a time-independent value, an expected signature of an equilibrium gel.
Similar conclusions have resulted from the experimental investigation of other clays, including high aspect ratio Montmorillonite~\cite{bohidar2015}.
In all cases, a region of thermodynamic instability has been detected at low densities, consistent with the theoretical expectation  by invoking an effective (although unknown) limited valence
of the clay particles.
It is interesting to note that in Laponite the progressive increase in the number of bonds is driven by an irreversible aggregation process.  In this respect, it could appear odd that Laponite gels
have been considered as an example of equilibrium gels.   This association has been supported by theoretical work~\cite{sciortino2011reversible,corezziJPCB,corezzi2012chemical} demonstrating that systems of limited valence particles --- when rapidly brought  to 
very small temperatures to simulate an irreversible aggregation process --- evolve in time
by progressively decreasing their effective temperature. Such a 
mapping between aging time and temperature (in equilibrium) has been invoked to interpret Laponite phase behavior  in the context of  limited-valence equilibrium phase diagram~\cite{ruzicka2011observation}.

\begin{figure}[h]
\begin{center}
\includegraphics[width=8cm]{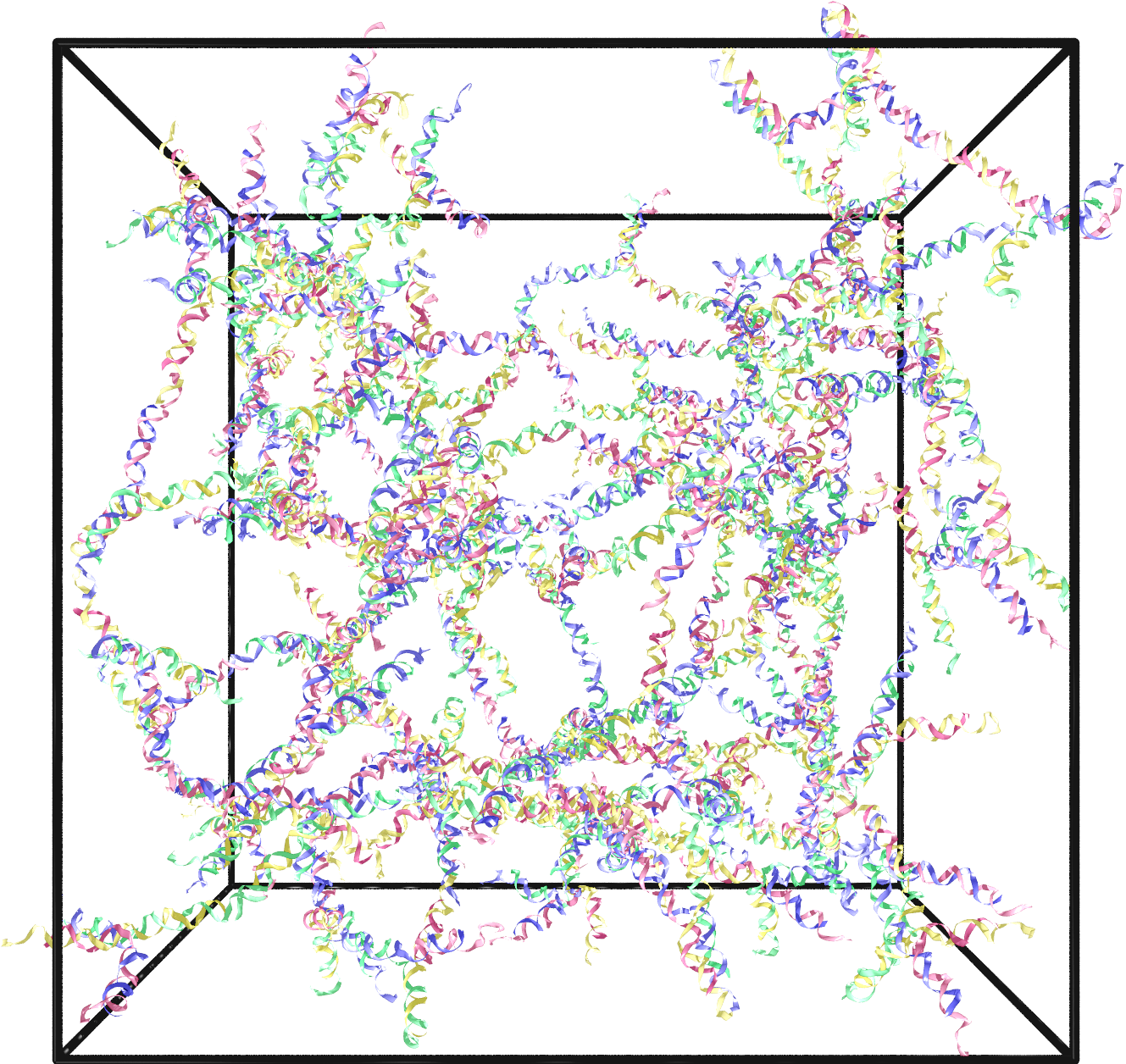}
\caption{Representation  of an all-DNA gel formed by DNA tetramers, adapted from Ref.\protect\cite{rovigatti2014accurate}.
Each nano star is formed by four DNA strands of 49 bases each. Binding between different nano-stars is provided by a
self-complememtary six-bases sequence located at the end of each arm. 
}
\label{fig:dnagel}
\end{center}
\end{figure}

\subsection{DNA-nanostars: phase diagram and equilibrium gels} A definitive experimental proof of the effect of the
valence on the phase-behavior requires  the ability to synthesise   bulk quantities of particles with 
different valence but identical  bond interactions.   This has been achieved capitalising on 
 recent developments in DNA nanotechnology,  i.e. on the idea that a proper selection of
DNA strands   can result into the self-assembly of 
constructs with well-defined shape.   Among these, DNA nanostars with a central
flexible core of unpaired basis and a  variable number of  double helix arms  have been realised
 (see Fig.~\ref{fig:example}).  The DNA nanostars self-assemble, starting from 
 a solution of properly selected isolated single strands at 90 C,  when slowly cooled below
 $T_{assembly} \approx 60$ C (a value slightly depending on the solution salt concentration), with a 
 high yield.  Below $T_{assembly}$, 
 these  constructs constitute stable ideal patchy colloidal particles with 
 well defined number of patches (the number of arms) and
 well defined interaction strength (controlled  by a six-base long  self-complementary sequence at the end of each arm).   The   $T$-dependence that characterizes DNA hybridization of the 
 six-base binding  sequence offers the opportunity to investigate the limited $T$-interval  between $T_{assembly}$
 and $T=0$ C, from the case of complete absence of inter-nano star bonds to the fully bonded 
 limit, i.e. the entire phase behaviour.

 The phase diagram  of tetravalent and trivalent DNA nanostars  (Fig.~\ref{fig:example}) has been
 experimentally investigated in Ref.~\cite{biffi_dna}.  For both valences, a clear phase-separation region
has been observed, with  the  critical $T$, the critical density  and the density of the coexisting dense phase  $\rho_l$
all decreasing on going from valence four to valence three, in full agreement with theoretical predictions~\cite{rovigatti2014accurate}. 
    
\begin{center}
\begin{figure*}[t]
\includegraphics[width=8 cm]{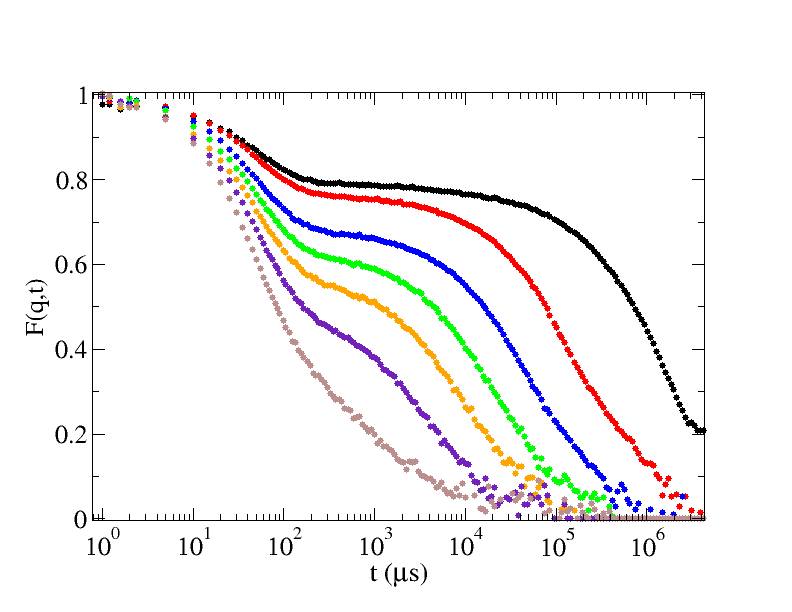}
\includegraphics[width=8 cm]{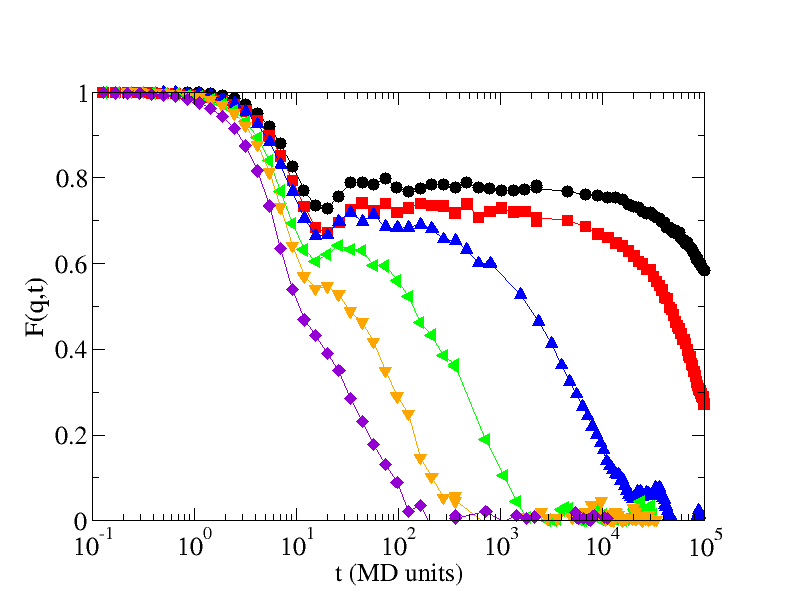}
\caption{Left: Density auto-correlation function $F(q,t)$ measured by Dynamic Light Scattering (DLS) for trivalent DNA nanostars at a concentration of $12$ mg/ml, with 150 mM added NaCl and various temperatures. From top to bottom: $T=40.1, 34.6, 32.2, 29.7, 27.5, 25.5, 18.1^{\circ} C$ (from Ref.~\cite{bomboi2015equilibrium}); (right) The same correlation function 
 calculated in Molecular Dynamics  simulation of  particles interacting via the $N_{\rm max}$\protect\cite{Zacca1,Zacca2} model,  with valence 3, at packing fraction $\phi=0.20$ and scaled temperatures, from top to bottom, $T^*=0.075, 0.08, 0.10, 0.125, 0.15, 0.17$.   In both systems, both the plateau height  and the relaxation time increase on cooling.}
\label{fig:corr}
\end{figure*}
\end{center}

The multi-valent DNA nanostars  form  equilibrium gels.  
Indeed  for $\rho > \rho_l$ the system  does not encounter phase-separation upon cooling,  and it 
remains  homogeneous for all $T$.    On cooling, different nanostars progressively bind to form a network 
where all possible bonds are formed (see Fig.~\ref{fig:dnagel}).
The correlation function of  the density fluctuations, measured via dynamic
light-scattering (DLS) experiments~\cite{biffi_dna, biffi2015equilibrium, bomboi2015equilibrium},  
shows a progressive increase both of the relaxation time $\tau_\alpha$  and of the relaxation amplitude on cooling. 
This is shown in Fig.~\ref{fig:corr}. The same figure also reports for comparison the density autocorrelation function for one of the limited valence models
previously investigated theoretically~\cite{Zacca1}.
 The increasing amplitude of the relaxation process on cooling is a signature of the progressive growth of  the   network.
 In the DNA-nanostar system, the relaxation time  increases by more than four orders of magnitude before
exceeding   the experimental accessible time window, a clear indication of the formation of an arrested state of matter on the experimental time scale.  The $T$-dependence of  $\tau_\alpha$ is well described by an Arrhenius law,
with an activation energy proportional to the binding free-energy of the sticky sequence, suggesting that
the decay of the density correlation is slaved to the elementary bond-breaking process.   It has also been reported
that $\tau_\alpha$ does not show a detectable wave-vector dependence, an uncommon behavior for a particle gel~\cite{biffi2015equilibrium}.   
An Arrhenius law also characterises  strong network-forming liquids, suggesting that equilibrium gels are indeed the colloidal analog of  these atomic and molecular systems. 

The DNA nanostar gel is clearly a very good candidate for experimentally testing the properties of equilibrium gels.  The viscoelastic properties of this class of materials will possibly be the focus of future investigation. Preliminary results suggesting that the viscosity also follows an Arrhenius law~\cite{biffi2015equilibrium}. The temperature and frequency dependence characterising the viscosity in equilibrium gels will hopefully be also investigated soon.

\subsection{Reentrant  equilibrium gels: a system with $T$-dependent valence.}  The ideas  previously discussed  are not only useful to interpret the phase behavior and 
the dynamics of particles with a fixed number of attractively interacting patches.  
They also apply to colloids with a  $T$-dependent valence, as telechelic star polymers~\cite{rovigatti2016soft} or
colloids interacting with dominant dipolar or quadrupolar potentials. In the last case  the
small valence builds up progressively on cooling due to the preferential sampling of the two or four
directions in which the interaction potential has a minimum. 
A typical example is provided by ferrofluids, i.e. colloidal solutions of particles with a 
net electric or magnetic dipole, promoting at low $T$ the formation of long polymer-like self-assembled
rings and chains of particles.   For larger $T$, branching points of different type~\cite{rovigatti2013branching}, effectively 
provide connections between different chains and rings. The decrease in  the number of 
branching points on cooling is effectively equivalent to  a reduction of the average valence of the system,  
which is accompanied by a progressive decrease of the  density of the coexisting liquid. 
One can thus expect that in dipolar systems, the gas-liquid phase separation, if present, 
shows a pinched shape, in which the density of the coexisting liquid progressively approaches the
density of the coexisting gas on cooling.  This idea, first proposed by  Tlusty and Safran for dipolar hard spheres~\cite{safrannew}, has indeed been realized in a model of  particles with dissimilar patches promoting respectively chaining and branching~\cite{prl-lisbona,lisbona_lungo}.
A pinched Safran-like phase diagram has also been observed in a simulation of a binary mixture of 
tetravalent and monovalent particles with competing bonding~\cite{roldan2013gelling,roldan2013phase}.  
Very recently, such a thermodynamic scenario has been reproduced experimentally with   tetravalent 
and monovalent DNA  nanoconstructs, encoding in the sequence of the DNA strands the physics 
of the competing bonding~\cite{bomboi2016re}.  On cooling, the system first assembles into a strong network formed by the  association of the tetravalent particles. On further cooling, the monovalent particles  preferentially start to bind
to the tetramers,  melting the gel and generating a fluid phase of  tetramers whose arm ends are 
unable to bind to other tetramers being capped by monomers.  The entire process, being based on
the low-valence equilibrium gel concept, is fully reversible.

\subsection{Thermodynamic stability of equilibrium gels.}  In principle, particle systems at low $T$
should  be metastable with respect to some ordered phase and slowly nucleate a crystal. Interestingly enough,
it has been shown that in limited valence systems, when bonds between particles are highly flexible, open crystals are strongly destabilized, so much that the liquid phase  becomes the thermodynamically stable phase~\cite{smallenburg2013liquids}. This suggests that equilibrium gels of  patchy particles can offer examples of thermodynamically stable equilibrium gels.   The  DNA nanostars  are a typical example of colloidal particles which can form
highly flexible bonds. Indeed, the four arms are highly mobile, due to the presence of unpaired bases
in the star center. It is thus not surprising that  numerical simulations based on the oxDNA model~\cite{ouldridge2013dna}, for the specific case of DNA nanostar gels, have confirmed the thermodynamic stability of the gel with respect to the crystal~\cite{rovigatti2014gels}. Other observations of flexibility-induced stability of the disordered phase have been reported in Ref.~\cite{prx}.  

\subsection{Biological applications.}  Proteins provide beautiful examples of patchy colloidal particles.  Hydrophobic patches or charged amminoacids~\cite{li2015charge}  often conspire to generate highly non-directional interaction potentials.  Proteins are often characterized by low critical volume fractions and by open crystals, indication of a significant directional component in the inter-protein interaction\cite{mcmanus2016physics}.
It is thus expected that the concept of equilibrium gel will be of relevance in biological applications. A preliminary indication in this direction can be found in the Ph.D. thesis of Jing Cai in the group of Alison M. Sweeney~\cite{tesi}. In this beautiful work, the squids cellular lens  is investigated to find out 
the biological origin of  the lens peculiar density gradient.  The results suggest  
that squids lens proteins  form a gel with a gradient in the gel concentration,  that correlates with a gradient in the valence  of the expressed proteins.

The physics of limited valence  has also been invoked to interpret the aggregation process in proteins induced  by the presence of  multivalent ions.   Interestingly, patches are activated by the presence of the ions,
allowing external control on the aggregation process.
The experimental investigation of Human Serum Albumine in the presence of YCl$_3$~\cite{zhang2012charge,zhang2014reentrant}  has shown the presence of a metastable liquid-liquid phase transition,  in which the coexisting liquid density progressively increases with the number of patches in the system, i.e. the salt concentration.  Theoretical modeling of the system via patchy particles consistently explains the experimental results~\cite{roosen2014ion}. Finally we note that patchy particle models 
can explain the experimental variation of the phase behavior for different mutants of the same protein~\cite{fusco2014characterizing} or for the modifications induced by fluorescent labelling~\cite{quinn2015fluorescent}.

\subsection{Other evidence}
Recent works by Dudukovic and Zukoski have reported the evidence of the formation of equilibrium molecular gels in a solution of Fmoc-diphenylalanine (Fmoc-FF) (see Fig.~\ref{fig:example})  in dimethyl sulfoxide. Upon addition of water or change of pH, the molecules self-assemble into fibers, which then form a space-filling fibrous network with large stiffness even at very low volume fractions\cite{dudukovic2014mechanical}. 
The molecules interact primarily via anisotropic reversible interactions, associated to strong $\pi-\pi$ stacking bonds of the side chains and hydrogen bonds.
The gel formation, occurring at sufficiently high $T$ in the absence of phase separation, has been interpreted in terms of reversible equilibrium gels~\cite{dudukovic2014evidence}, in analogy with patchy particle systems.
Finally, a series of works by Odriozola and coworkers~\cite{meneses2013towards,varga2013empty,varga2014empty} have investigated the role of the shape in attractive particles, by examining the phase behavior of attractive
ellipsoids (see Fig.~\ref{fig:example}). Even if the particles interact with a simple square-well potential, the anisotropic shape restricts the formation of possible bonds, providing limited valence to the particles. Extensive Monte Carlo simulations have shown that the gas-liquid critical point moves towards low particle densities in full analogy with patchy spheres~\cite{varga2013empty}. Furthermore, for oblate particles, also the critical temperature extrapolates to zero as in patchy spheres,  while in prolate ones it seems to approach a finite value. The variation of the critical point is accompanied by a shrinking of the phase coexistence region and consequently the existence of a liquid phase at very low particle densities. This opens up the possibility to observe equilibrium gels 
in this new class of systems. Recent calculations suggest that  equilibrium gel formation should be facilitated 
 for rod-like rather than disk-like attractive ellipsoids~\cite{varga2014empty}.

\section{Percolation at gelation}
Some systems show gelation in equilibrium crossing (or close to) the percolation line.  Despite these systems are not equilibrium gels of limited valence particles, they share with them several
interesting features.

A very interesting gel-forming system of this type, first studied by Appell and co-workers~\cite{appell,filali2004robust}, is composed by microemulsion droplets in solution with telechelic polymers.   In this system the polymer core prefers  to be surrounded by the solvent, while the two  ends  preferentially explore the interior of the microemulsion. When the two ends are located in different droplets, they effectively provide 
a transient link between them.  The fraction of telechelic polymers in solution thus controls the effective average valence. 
Interestingly, in this system gelation can coincide with percolation.  Indeed the lifetime of the inter-droplets bond is  controlled by the average residence time of the polymer end inside the microemulsion. As a result, the  transient network gel restructures itself on  the same time scale as the bond lifetime.  By tuning the residence time (a quantity that is completely decoupled from the number of inter-droplets bonds) it is thus possible
to control the location of the gelation line~\cite{hurtado}. For very long residence times, gelation coincides with percolation~\cite{hurtado}.
Such a system  provides an interesting example of decoupling the bond lifetime from the strength of the interparticle interaction~\cite{delgado2,saikabond}.
The density autocorrelation functions measured by DLS show an increasing plateau on increasing the concentration of
 telechelic polymers~\cite{appell}), as in the DNA nanostars systems.  But differently from the DNA nanostar case,
 the $\alpha$-relaxation time does not depend on the number of bonds, a clear evidence of the
 separation of the bond lifetime from the thermodynamics controlling the aggregation of the microemulsion droplets. Interestingly, 
 the $\alpha$-relaxation time is wavevector independent~\cite{appell},  similarly to the DNA nanostar case.

Another  system in which dynamic arrest and gel formation appear to be related to the crossing of the percolation line is 
provided by octadecyl-coated silica particles suspended in n-tetradecane~\cite{eberle2011dynamical,eberle2012dynamical}.  In this system 
the inter-particle potential is tuned by temperature.  The $T$ affects the brush conformation leading to a liquid-to-solid phase
transition of the brush itself.  Static scattering experiments suggest that the inter-particle potential can be rather well approximated
with a square-well potential of width 1\% of the particle diameter~\cite{eberle2011dynamical}.  
However, accurate event-driven simulations~\cite{foffi2005scaling} have shown that for this theoretical model the bond lifetime along the percolation locus is too fast to generate gelation. The disagreement between numerical and experimental studies suggests that in this system the conformational transition of the brush may at the origin of the large bond lifetime at percolation.

\section{Conclusions and future directions}

We expect a significant activity focused on equilibrium gels in the next years. Several interesting  problems need to be tackled. 
To start with, an effort is requested to move from the design of single particles to the  production of bulk quantities  of them.
Indeed, while a vast zoo of novel patchy colloids and colloidal molecules  has been reported in recent years,  scalability in their synthesis 
and bulk production has been rarely achieved.  
Only production of large  quantities of  limited-valence particles, with with designed and controlled  connectivity, bond strength and bond lifetime,
 will make it possible   
--- in addition to the few experimental systems studied so far --- to test 
  the predictions of theoretical and numerical investigations and
to study in full details the self-assembly process and  the collective response  of equilibrium gels.  At the same time, it will be possible
to clarify the analogies and differences between equilibrium gels of limited valence particles and physical gels  in which gelation
is associated to crossing of some sort of a percolation line.

We also foresee a significant effort towards deepening the connection between patchy colloidal particles  and proteins. 
More specifically, we need to quantify the protein-protein 
effective interaction and its patchiness.  This will also require understanding cases in which  isotropic interactions coexist with more directional ones~\cite{KumarJCP07}, still retaining the 
windows of densities where equilibrium gel forms. 

Concerning the dynamics of equilibrium gels of limited valence particles, several  questions are still open:

 Is there a peculiar dynamics associated to the crossing of the glass and equilibrium gel lines~\cite{krzakala2008lattice,Zacca2} ?  Is the competition between arrest due to bonding and arrest due to excluded volume caging capable to produce anomalous dynamics\cite{Zacca2} as in the case of short-range attractive colloids\cite{a4}?
 
Bond flexibility has recently been shown to have a large impact on  the relative thermodynamic stability of equilibrium gels compared  to the one of
crystals~\cite{smallenburg2013liquids}.  Has flexibility a similar role also on the network restructuring dynamics and hence on its mechanical properties ?

A wave vector independent relaxation time $\tau_{\alpha}$ characterise the slow decay of the density fluctuations in DNA-gels~\cite{biffi2015equilibrium} and  microemulsion droplets linked by telechelic polymers~\cite{appell}.  What is the microscopic mechanism controlling such length-scale invariant process ?  Does it have a counterpart in atomic and molecular network-forming liquids ? How is the bond lifetime related to  $\tau_{\alpha}$ and to the system viscosity ?
What are the microscopic mechanisms controlling density and stress fluctuations  in these materials ?

We are confident that  these questions will be answered in the near future.

\subsection{Acknowledgement}
We thank A. M. Sweeney, L. Rovigatti and N. Gnan for useful discussions. FS thanks T. Bellini, S. Biffi and R. Cerbino for the ongoing scientific collaboration on DNA gels. FS and EZ acknowledge support from ETN-COLLDENSE (H2020-MCSA-ITN-2014, Grant No. 642774). 
EZ acknowledges support from ERC MIMIC (ERC-CoG-2015, Grant No. 681597).

\bibliography{tetra}

\end{document}